\documentclass[reprint,amssymb, amsmath, aip,cha,twocolumn]{revtex4-1}
\usepackage{graphicx}
\usepackage{color}
\usepackage{epsfig}
\usepackage{amssymb}
\usepackage{bm}%
\usepackage{epstopdf}
\usepackage[colorlinks=true,linkcolor=blue]{hyperref}%
\expandafter\ifx\csname package@font\endcsname\relax\else
 \expandafter\expandafter
 \expandafter\usepackage
 \expandafter\expandafter
 \expandafter{\csname package@font\endcsname}%
\fi
\begin{document}
\title{Dust density waves in a dc flowing complex plasma with discharge polarity reversal}%

\author{S. Jaiswal}
\thanks{Presently at Physics Department, Auburn University, Auburn, AL 36849 USA}
\author{M. Y. Pustylnik}
\email{mikhail.pustylnik@dlr.de}
\author{S. Zhdanov}
\author{H.M. Thomas}
\affiliation{Institut f\"ur Materialphysik im Weltraum, Deutsches Zentrum f{\"u}r Luft- und Raumfahrt (DLR), 82234 We{\ss}ling, Germany.}%

\author{A.M. Lipaev}
\author{A.D. Usachev}
\author{V.I. Molotkov}
\author{V.E. Fortov}
\affiliation{Joint Institute for High Temperatures, Russian Academy of Sciences, 125412 Moscow, Russia}

\author{M.H. Thoma}
\affiliation{I. Physikalisches Institut, Justus-Liebig-Universit\"at Gie\ss en, 35392 Gie\ss en, Germany}

\author{O.V. Novitskii}
\affiliation{Gagarin Research and Test Cosmonaut Training Center, 141160 Star City, Moscow Region, Russia}

\date{\today}
\begin{abstract}
We report on the observation of the self-excited dust density waves in the dc discharge complex plasma.
The experiments were performed under microgravity conditions in the \mbox{Plasmakristall-4} facility on board the International Space Station.
In the experiment, the microparticle cloud was first trapped in an inductively coupled plasma, then released to drift for some seconds in a dc discharge with constant current.
After that the discharge polarity was reversed.
DC plasma containing a drifting microparticle cloud was found to be strongly non-uniform in terms of microparticle drift velocity and plasma emission in accord with [Zobnin {\it et.al.}, Phys. Plasmas {\bf 25}, 033702 (2018)].
In addition to that, non-uniformity in the self-excited wave pattern was observed: In the front edge of the microparticle cloud (defined as head), the waves had larger phase velocity than in the rear edge (defined as tail).
Also, after the polarity reversal, the wave pattern exhibited several bifurcations: Between each of the two old wave crests, a new wave crest has formed.
These bifurcations, however, occurred only in the head of the microparticle cloud.
We show that spatial variations of electric field inside the drifting cloud play an important role in the formation of the wave pattern.
Comparison of the theoretical estimations and measurements demonstrate the significant impact of the electric field on the phase velocity of the wave.
The same theoretical approach applied to the instability growth rate, showed agreement between estimated and measured values.
\end{abstract}
\maketitle
\section{Introduction}\label{sec:intro}
Dust density waves (or dust acoustic waves) predicted theoretically by Rao, Shukla and Yu (RSY)\cite{dawrsy} and observed experimentally for the first time \cite{barkan_merlino} quite long ago remain a significant research topic in dusty plasma physics also in the present time\cite{merlino25}.
The reason for that is that self-excitation of these waves is observed in all types of discharges,~\textendash~dc\cite{khrapakag_waves, ethomas_waves}, inductively coupled rf\cite{usachev_waves}, capacitively coupled rf\cite{pkenefedov_waves, piel_waves},~\textendash~under laboratory\cite{schwabe_highlyreswaves} and microgravity\cite{piel_clustering} conditions, with micrometer-\cite{nosenko_qmachwaves} and nanometer-sized\cite{kortshagen_waves} dust particles.
\par
Ion-streaming instability associated with sheath or bulk electric field is usually identified as a self-excitation mechanism of the dust density waves \cite{rosenberg_sewaves, khrapakag_waves}.
Although, so far, there seem to be no experimental observations that contradict to this hypothesis, on the other hand, to the best of our knowledge, there were no experiments conducted that would directly target the dependence of the self-excited wave patterns on the fundamental parameter which controls the self-excitation mechanism, namely, on the electric field.
\par
Self-excitation of dust density waves is one of the aspects of dust-plasma interactions.
It is, therefore, of interest, for technological plasma processing: it has been, e.g., shown that dust density waves enhance agglomeration of particles in plasmas\cite{du_wagglomeration}.
From the point of view of complex plasma research, which uses suspensions of microparticles to model classical phenomena of condensed matter physics\cite{cprmp}, self-excited density waves represent an unwanted plasma effect, which becomes an obstacle in creating calm 3D suspensions at low neutral gas pressure.
In addition to that, dispersion relations for the dust density waves are already used as diagnostic tools to reconstruct the information on the local plasma parameters from the self-excited wave pattern \cite{benjamin_daw_pop, benjamin_daw_pre}.
Therefore, deeper experimental investigation of the self-excitation of the dust density waves is required.
\par
Plasmakristall-4 (\mbox{PK-4}) facility on board the International Space Station\cite{pk4rsi} is the best-suited setup for studying self-excited density waves in complex plasmas.
Its versatile power supply controlling the dc discharge allows to vary the average electric field experienced by the microparticles, while keeping other parameters practically constant.
In this paper, we present the first observations of the effect of electric field manipulation (namely, of the discharge polarity reversal), on the self-excited wave pattern.
 \par
The paper is organized as follows. In the next section (Sec.~\ref{sec:setup}), we present the experimental set-up and describe the experimental procedure. 
In Sec.~\ref{sec:results}, we show and discuss the experimental results and in Sec.~\ref{Sec:Conc}, we draw the conclusions. 
\section{Experimental setup and procedure}\label{sec:setup}
\begin{figure}[!ht]
\includegraphics[scale=0.5]{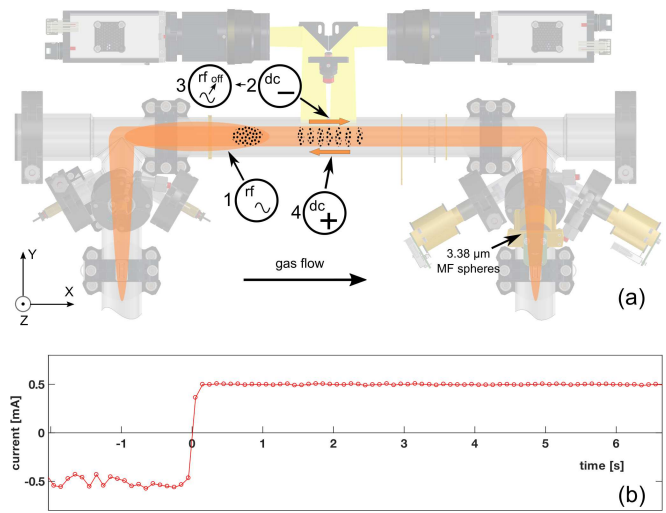}
\caption{\label{fig:exp_proc}(a) Schematic representation of the experimental procedure. 
Initially (step 1), the microparticles are trapped in the rf discharge. 
In step 2, the dc discharge with the current $-0.5$~mA is switched on and right after that (step 3), the rf discharge is switched off. 
In step 4, the dc discharge polarity is reversed. During both discharge polarities, the dust density waves are excited. 
The arrows shows the direction of motion of the microparticle cloud.
(b) Evolution of the discharge current during the experiment.
Time intervals corresponding to negative and positive current values represent step 2 and step 4 of the procdure, respectively.}
\end{figure}

\begin{figure*}[!ht]
\centering
  \includegraphics[width=1\textwidth]{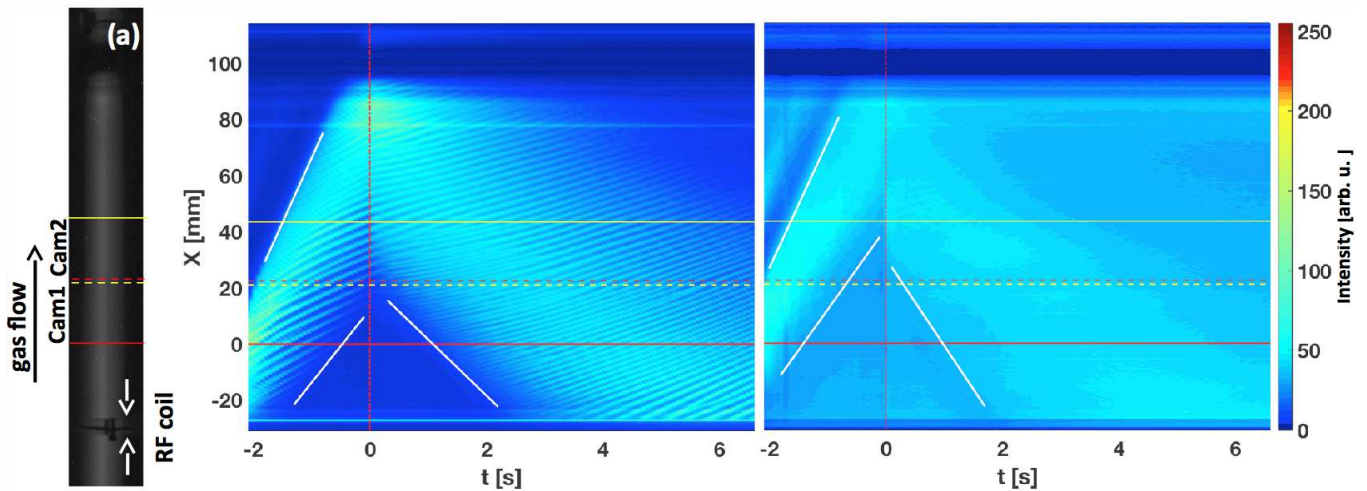}
  \caption{\label{fig:global_view_waves} (a) Image of the microparticle-free dc discharge taken through the $585.2$~nm channel of the plasma glow observation (PGO) system (see Section~\ref{sec:setup}). 
Spatiotemporal patterns (b) of the microparticle dynamics obtained through the neutral density channel of the PGO system and (c) of the plasma emission dynamics obtained through the $585.2$~nm channel of the PGO system. 
Horizontal solid and dashed lines show, respectively, the outer and inner boundaries of the FoVs of two particle observation (PO) cameras: Red lines correspond to PO camera 1 and yellow lines~\textendash~to PO camera 2. 
Vertical red dash-dotted lines correspond to the moment of the discharge polarity reversal. 
The white lines show the slopes corresponding to the velocities of the head and tail edges of the drifting cloud.}
\end{figure*}

%
In the \mbox{PK-4} facility \cite{pk4rsi}, the plasma is created in a cylindrical (3 cm inner diameter and 20 cm working area length) glass chamber by means of a dc discharge (Fig.~\ref{fig:exp_proc}(a)). 
Neon at 40 Pa pressure was used as a working gas. 
The pressure was controlled by a pressure controller.
The valve of the flow controller was kept closed during the experiment to minimize the gas flow through the plasma chamber, however, a small gas leak of about 0.1 sccm was still observed.
Monodisperse plastic (melamine formaldehyde) microspheres with a diameter of $d= 3.38\pm0.07~\mu$m and a mass of $m_d=3.1\times10^{-11}~$g from a shaker dispenser were used. 
Inductively-coupled plasma with $0.4$~W forward power produced by a movable rf coil (location shown in Fig.~\ref{fig:global_view_waves}(a)) was used to trap them. 
The microparticles were illuminated by a sheet of a green laser. 
The three-channel kaleidoscopic plasma glow observation (PGO) system\cite{pk4rsi} was used to monitor the microparticle (neutral density channel) as well as plasma ($585.2$~nm, Ne\,I spectral line channel) dynamics in the entire working area at $15$~fps frame rate and $30$~ms exposure time and a resolution of $430~\mu$m/pixel. 
Two particle observation (PO) videocameras (recording at 70 fps) positioned as shown in Fig.~\ref{fig:global_view_waves}(a) were registering the dynamics of microparticles only.
The total size of the FoV of both PO cameras was approximately $44\times7$~mm$^2$ with the resolution of about $14~\mu$m/pix.
Experiments were conducted in the flight model of \mbox{PK-4} under microgravity conditions on board the International Space Station.
\par
Before starting the experiment, the microparticles have to be injected into the plasma, transported into the working area and then trapped there. 
Microparticle transport and trapping techniques are in detail described in Ref.~\onlinecite{pk4rsi}. 
In case of this particular experiment, a combined transport and trapping technique was used, in which transport was performed with the help of a dc discharge and trapping with the help of an inductively coupled rf discharge.
\par
The experimental procedure is sketched in Fig.~\ref{fig:exp_proc}.
After trapping the microparticles by the rf plasma, the dc discharge was switched on and rf power was simultaneously switched off. 
A constant current of -0.5 mA was kept for 2 s and then the polarity of the discharge was reversed. 
The constant current of 0.5 mA was kept during 8 s. 
The evolution of the discharge current during the experiment is shown in Fig.~\ref{fig:exp_proc}(b). 
\par
Figs.~\ref{fig:global_view_waves}(b) and \ref{fig:global_view_waves}(c) represent the spatiotemporal patterns of the light intensity in the neutral density and $585.2$~nm channels of the PGO system, respectively.
The patterns were obtained by averaging the intensities of 10 central pixels over the direction of $Y$ axis for every frame in the respective channels.
As follows from those two figures, switching on the dc current leads to the drift of the microparticle cloud and generation of the self-excited density waves inside it.
Plasma emission follows the dynamics of microparticles.
\par
Plasma parameters in the absence of microparticles were calculated using empirical formulae given in Ref.~\onlinecite{pk4rsi}. 
For a specific discharge condition (discharge current $\left| I_{dc} \right| = 0.5~$mA, and pressure $P = 40$~Pa), the electron density, electron temperature, and axial electric field strength were estimated to be:  $n_e=1.27\times10^8~$cm$^{-3}$, $T_e=8.56~$eV and $E=2.4~$V/cm, respectively. 
The number density of the microparticles ($n_d$) was estimated to be $\sim 10^5~$cm$^{-3}$. 

\section{Results and discussion}\label{sec:results}
\subsection{Drift of the microparticle cloud}
\begin{table}
\caption{\label{tab:table1}Drift velocities (in mm/s) of the edges of the microparticle cloud measured from Fig.~\ref{fig:global_view_waves}.}
\begin{ruledtabular}
\begin{tabular}{ l  c  c  c }

										& Edge	&$ I_{dc}=$ 	&$I_{dc}=$\\
										&		&$-0.5$~mA	&$0.5$~mA\\
\hline
Microparticles (Fig.~\ref{fig:global_view_waves}(b))	& head 	& $46\pm1$  	& $-19\pm1$\\
									 	& tail 	& $26\pm1$ 	&		    \\
Plasma glow (Fig.~\ref{fig:global_view_waves}(c))	& head 	& $45\pm1$  	& $-31\pm1$\\
										&tail		& $29\pm1$
\end{tabular}
\end{ruledtabular}
\end{table}
As it was already mentioned, the microparticle cloud starts drifting in a dc discharge. 
The direction of the drift is determined by the direction of the electrostatic force acting on the negatively-charged microparticles.
We define the front edge of a drifting cloud as its {\it head}, and the rear edge of the cloud as its {\it tail}.
\par
Fig.~\ref{fig:global_view_waves}(b) clearly shows, that the microparticle cloud does not exhibit a rigid drift: Its head moved significantly faster compared to its tail.
The cloud, therefore, stretches along the discharge axis.
The velocities of the head and tail are with a good accuracy constant in time.
Their values are summarized in Table~\ref{tab:table1}.

\subsection{Wave pattern}
\label{sec:wpattern}
The generation of self-excited dust density waves can be seen in a PGO space-time diagram (Fig.~\ref{fig:global_view_waves}(b)).
From this figure, it is already obvious that the waves do not change the propagation direction after the polarity reversal.
However, this figure does not allow to observe the details of the wave propagation due to the limited resolution of the PGO system.
\par
\begin{figure}[!ht]
\includegraphics[width=0.4\textwidth]{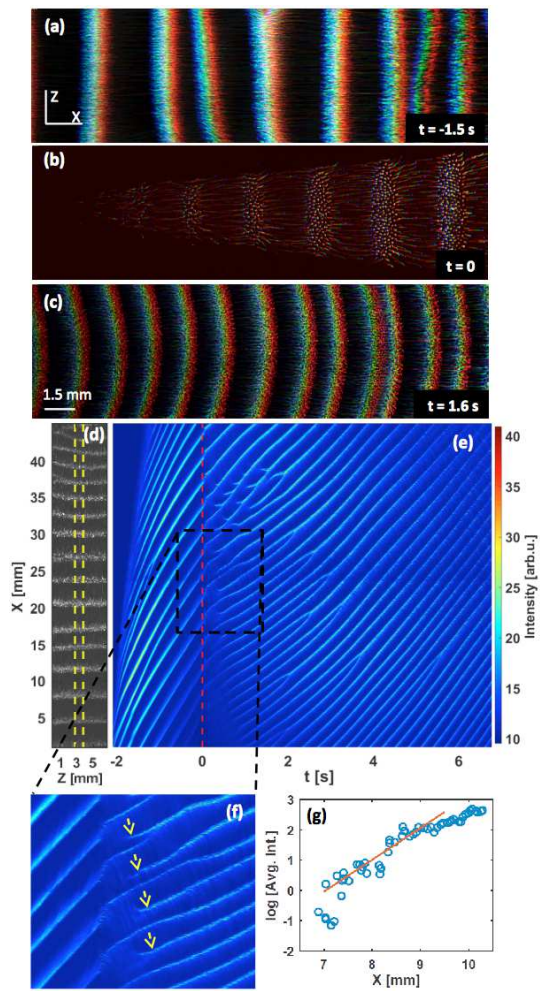}
\caption{\label{fig:wave_PO_camera}Color-coded Images of self-excited waves recorded by PO cameras (a) before, (b) during, and (c) after polarity reversal. 
Several consecutive raw images were superimposed to obtain one color-coded image. 
Blue color corresponds to the earliest raw image whereas red color~\textendash~to the latest one.
(d) A single image of the density wave, combined from respective PO camera~1 and PO camera~2 images.
Region between the yellow dashed lines was used to create a spatiotemporal pattern of the intensity.
(e) Spatiotemporal pattern of the intensity constructed using the images, combined from respective PO camera~1 and PO camera~2 images.
Vertical dashed line shows the moment of polarity reversal.
Direction of the wave propagation does not change after the polarity reversal.
Waves exhibit large amplitude and bifurcations, leading either to the birth or death of the wave crests.
(f) A zoom into the spatiotemporal pattern demonstrating the bifurcations, leading to birth of a new wave crests (shown with arrows) after discharge polarity reversal. 
The new crest appears between the two existing crests. 
(g) Evolution of the amplitude of a newborn wave crest. 
Circles correspond to the measured amplitudes and the red line represents the exponential fit to the rapid growth phase.}
\end{figure}

Significantly higher resolution is provided by the PO cameras.
Images in Figs.~\ref{fig:wave_PO_camera}\mbox{(a)-(c)} show the dynamics of wave crests before, around and after the polarity reversal, respectively. 
Sequences of eight consecutive raw images from PO camera~1 were superimposed  to obtain each of the three images.
Color-coding in Figs.~\ref{fig:wave_PO_camera}\mbox{(a)-(c)} reflects time.
Wave crests propagating in the same direction are observed before and after the polarity reversal.
Around the polarity reversal, the microparticles loose their drift velocity, but preserve the spatial oscillatory structure of their number density.
In Fig.~\ref{fig:wave_PO_camera}(d), a combination of the images from PO camera~1 and PO camera~2 is shown.
Such images were used to produce a spatiotemporal pattern (Fig.~\ref{fig:wave_PO_camera}(e)) by averaging the pixel intensities over the $Z$ direction across narrow stripes (shown in  Fig.~\ref{fig:wave_PO_camera}(d)).
\par
After the polarity reversal, the wave pattern exhibits several bifurcations.
The wave crests arriving at the polarity reversal continue their propagation further.
However, in the head of the microparticle cloud, new crests are formed between each of the two old ones (Fig.~\ref{fig:wave_PO_camera}(f)).
A similar phenomenon was observed in Ref.~\onlinecite{chengran_epl} at the interface separating the two parts of a microparticle cloud consisting of microparticles of two different sizes.
The new crests propagate for about $1.5-2$~s and then merge with the old ones.
Along a wave crest crossing the polarity reversal line at $X\approx30$~mm, three newborn crests are observed, which, however, merge with the old crest within a fraction of a second.
Further (at $X>40$~mm) inner wave crests do not exhibit any bifurcations at all.
\par

Fig.~\ref{fig:wave_PO_camera}(g) presents the evolution of the amplitude of a wave crest born as a consequence of polarity reversal.
After birth, its intensity grows exponentially with a typical rate of $1.0\pm0.2$~mm$^{-1}$ and then continues increasing at a slower rate.
\par
\begin{figure}[!ht]
\includegraphics[width=0.4\textwidth]{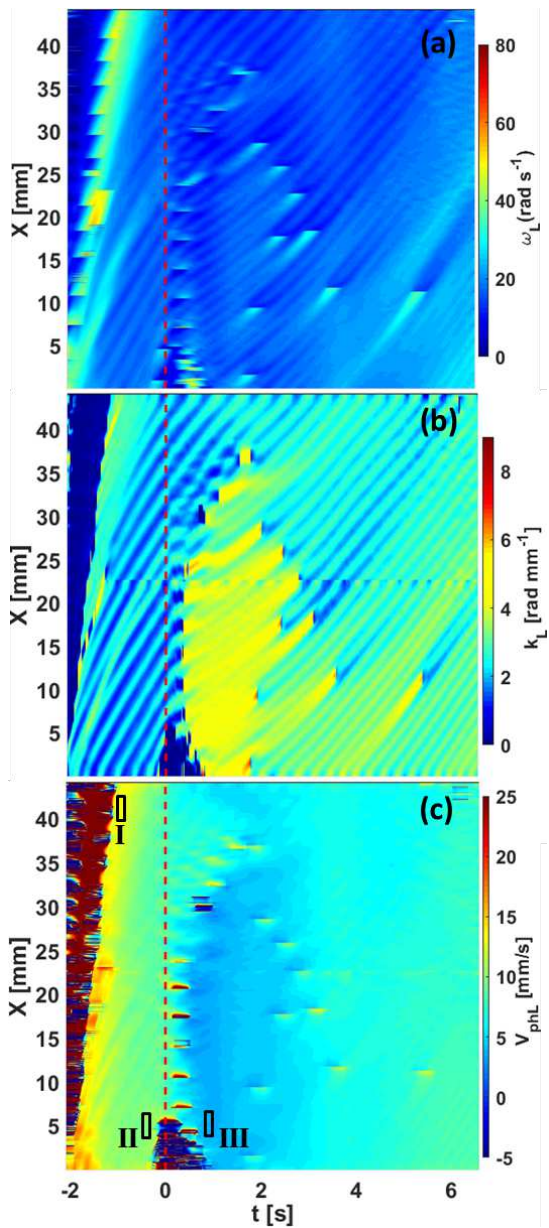}
\caption{\label{fig:hilbert_result}Spatiotemporal patterns of the (a) frequency (b) wave number and (c) phase velocity obtained using the Hilbert transformation from the spatiotemporal pattern of the image intensity in Fig.~\ref{fig:wave_PO_camera}(e). 
All three quantities are measured in the reference frame of the \mbox{PK-4} cameras.
The red dashed line shows the time of discharge polarity reversal.
Black rectangles mark three areas used for measurement of the wave phase velocity with respect to the cameras (see Table~\ref{tab:table2}).}
\end{figure}

To investigate the local characteristics of the observed wave pattern, the space-time diagram in Fig.~\ref{fig:wave_PO_camera}(e) was processed using the Hilbert transformation.
This technique has already been applied to study waves in complex plasmas \cite{menzel_hilbert_pre,jeremiah1_pre}.
For that, a complex function $J(X,t)$ was constructed out of the image intensity $I(X,t)$ (which is proportional to the local microparticle density $n_d(X,t)$):
\begin{align}
J(X,t)&=I(X,t)+i\hat{I}(X,t)\\\nonumber
        &=A(X,t)\text{exp}(i\phi(X,t)),\nonumber
\end{align}
where $\hat{I}(X,t)$ is the Hilbert transform of $I(X,t)$, $A(X,t)$ and $\phi(X,t)$ are the local values of the wave amplitude and phase, respectively.
Local values of frequency $\omega_L(X,t)$ and wave number $\text{k}_L(X,t)$ can be obtained by differentiating phase over time and space, respectively: $\omega_L=\partial \phi /\partial t$ and $\text{k}_L=-\partial \phi/\partial X$.
The resulting plots are shown in Figs.~\ref{fig:hilbert_result}(a) and (b).
Stripes stretching along the wave crests mean variation of $\omega_L(X,t)$ and $\text{k}_L(X,t)$ within the wave period, which suggests non-linearity of the waves (indicated also by large amplitude of the waves seen in Fig.~\ref{fig:wave_PO_camera}).
Regions where the values undergo rapid changes correspond to bifurcations~\textendash~either to the birth of wave crests or to their merging.
\par

Local phase velocity of the wave can be defined as $\text{V}_{phL}=\omega_L/\text{k}_L$.
The respective plot is shown in Figs.~\ref{fig:hilbert_result}(c).
Spatiotemporal pattern $\text{V}_{phL}(X,t)$~exhibits much less stripes than $\omega_L(X,t)$ and $\text{k}_L(X,t)$ and small outstanding islands correspond to bifurcations.
Note that all the three values $\omega_L(X,t)$, $\text{k}_L(X,t)$ and $\text{V}_{phL}(X,t)$ are measured in the reference frame of PK-4 cameras, which are at rest, whereas the medium in which the wave propagates is moving.
$\text{V}_{phL}$, therefore, cannot be directly related to the plasma parameters.
\begin{table}
\caption{\label{tab:table2}Phase velocities of the density waves determined for the three regions marked in Fig.~\ref{fig:hilbert_result}(c).}
\begin{ruledtabular}
\begin{tabular}{lccr}
Reference frame	&	& Velocity (mm/s) & \\
				& Region I 		& Region II 	& Region III\\
\hline
Cameras 				& $13.0\pm0.5$ 	& $10\pm0.2$ 	& $3.8\pm0.1$\\
Drifting medium			& $-33\pm2$ 		& $-16\pm1$	& $23\pm1$\\
\end{tabular}
\end{ruledtabular}
\end{table}

\subsection{Plasma non-uniformity}
As already mentioned above, the head of the microparticle cloud drifting in a dc discharge moves faster than its tail (see Fig.~\ref{fig:global_view_waves}(b) and Table~\ref{tab:table1}).
Plasma emission follows the motion of the micropaticles to some extent, as can be seen in Fig.~\ref{fig:global_view_waves}(c).
Before the polarity reversal, emission of the plasma is stronger in the head of the microparticle cloud (in accord with Ref.~\onlinecite{Zobnin2018}, also the striation ahead of the drifting cloud is visible).  
At the moment of polarity reversal, a sudden increase of the emission intensity at the edge of the cloud (which was a tail before polarity reversal and becomes a head after) is observed.
Plasma non-uniformity reveals itself also in the wave pattern.
Namely, the bifurcations leading to the birth of the new wave crests tend to appear in the head of the drifting cloud after the polarity reversal.
\par
Therefore, in our experiment, a dc plasma containing a drifting microparticle cloud was found to be non-uniform in three observed parameters, namely drift velocity of the microparticles, plasma emission and pattern of the self-excited density waves.

\subsection{Force balance}
Although the microparticle cloud does not move rigidly in a dc discharge, their motion can be described by means of average drift velocities, obtained by tracing the peak of plasma emission in Fig.~\ref{fig:global_view_waves}(c).
This brings us to the following two equations:
\begin{eqnarray}
\eta m_d V_{-} &=& |Z_deE|+F_{gas}\label{eq:drift_1},\label{Eq:ForceB1}\\ 
-\eta m_d V_{+} &=& -|Z_deE|+F_{gas}\label{eq:drift_2}.\label{Eq:ForceB2}
\end{eqnarray}    
Here, $V_{-}$ and $V_{+}$ are the average drift velocities for negative and positive discharge polarities, respectively, $\eta\approx 105$~s$^{-1}$ under our experimental conditions, is the momentum transfer rate in dust-neutral collisions, and 
\begin{equation}
F_{gas} = \eta m_d V_{gas},
\end{equation}  
where $V_{gas}$ is the gas flow velocity.
We have neglected the ion drag force for simplicity.
Since the temperature sensors at the edges of the working area of the plasma chamber exhibit no measurable temperature difference, we have neglected also the thermophoretic force.
We have measured $V_{-}=34$~mm/s and $V_{+}=20$~mm/s.
Solving Eqs.~\ref{eq:drift_1} ~and~\ref{eq:drift_2}, results in $V_{gas}=\left(V_{-}-V_{+}\right)/2=7$~mm/s, which is in good agreement with the gas flow velocity estimated from the measured gas flow rate of $0.1$~sccm.
Therefore, it is the gas flow which makes the waves propagate in the same direction (with respect to the cameras) irrespective of the polarity of the dc discharge.
Using the microparticle-free value of the axial electric field, we can estimate $Z_d\approx2200$. 

\subsection{Wave phase velocities}
The phase velocities presented in Fig.~\ref{fig:hilbert_result}(c) are measured with respect to \mbox{PK-4} cameras.
To be able to treat the phase velocities as local properties of the medium, one needs to transfer the entire Fig.~\ref{fig:hilbert_result}(c) into the reference frame associated with the drifting microparticle cloud.
This requires knowledge of average drift velocity at every time-space point.
Due to high velocities of the microparticles and long tracks they leave in the camera images, it was impossible to trace particles and measure the drift velocities everywhere inside the cloud directly.
\par
However, the flow velocities are reliably determined at the edges of the cloud (Table~\ref{tab:table1}).
The ``true'' phase velocities in the three regions marked in Fig.~\ref{fig:hilbert_result}(c) are defined as
\begin{equation}
V_{ph} = V_{phL}-V,
\end{equation}  
where $V$ is the local drift velocity of the microparticle cloud.
$V$ can be taken equal to the drift velocity of the respective edge, whereas $V_{phL}$ is determined by averaging over the respective region in Fig.~\ref{fig:hilbert_result}(c).
The results are summarized in Table~\ref{tab:table2}.
Obviously, the values of the phase velocity are quite (up to factor of~$2$) different in the head and in the tail of a drifting cloud.

\subsection{Acoustic velocity of dust density waves in the presence of the electric field}
As mentioned above, the physical mechanism of dust density waves self-excitation is the streaming instability due to the ion flow, i.e. it essentially involves the electric field.
Very often \cite{LADAW, fink_autowaves_epl}, an estimation of the dust acoustic velocity using RSY approach \cite{dawrsy} for the uniform and isotropic plasma {\it without} an electric field already gives a reasonable agreement with the observed phase velocities.
In principle, this is also true for the waves that we observe in \mbox{PK-4}.
For a dense microparticle cloud where most of the negative charge sits on microparticles (since in our case, $Z_dn_d$ is larger than the unperturbed ion density, we assume $Z_dn_d\approx n_i$) dust acoustic velocity \cite{fink_autowaves_epl}
\begin{equation}
C_{RSY}=\omega_{pd}r_{di}=\sqrt{Z_d\frac{kT_i}{m_d}}, \label{eq:cda}
\end{equation}
where $\omega_{pd}=\sqrt{Z^2_de^2n_d/m_d\epsilon_0}$ and $r_{di}=\sqrt{\epsilon_0kT_i/n_ie^2}$ are dust plasma frequency and ion Debye radius, respectively, and $T_i$ is the ion temperature.
Taking $T_i=0.025$~eV and $Z_d=2200$, yields $C_{RSY}=17.1$~mm/s, which is already of the order of the values in Table~\ref{tab:table2}. 
\par
According to Eq.~\ref{eq:cda}, $C_{RSY}$ is (explicitly) independent of the microparticle number density, and the only parameter, which may be responsible for the variation of the phase velocity is the microparticle charge $Z_d$ (we do not expect variation of either $T_i$ or $m_d$ within the space-time domain of our experiment).
However, according to the simulations in Ref.~\onlinecite{Zobnin2018}, $Z_d$ in the head and in the tail are very close to each other since they approach the charge of a single particle in a dc discharge.
Even taking into account the entire amplitude of the variation of $Z_d$ inside a drifting cloud, it will be difficult to explain the observed variation of the phase velocity.
Eq.~\ref{eq:cda}, therefore, seems to be appropriate for the order of magnitude estimation of the phase velocity of the dust density waves, however, it turns out to be insufficient to qualitatively explain the observed spatiotemporal trends.
\par
The physical reason for that is not only the formal presence of electric field.
RSY-waves, if excited in our plasma, would be damped within the time of the order of $1/\eta$, and would therefore never reach the observed high amplitudes.
The pattern, we observe, is essentially a dissipative structure, which is fed with energy by the ion streaming instability and dissipates energy due to neutral drag.
Therefore, we adopted a different approach to the velocity of the dust density waves in complex plasmas with electric field.
In a dispersion relation from Ref.~\onlinecite{usachev_waves} we limited the Taylor expansion on $\omega/\omega_{pd}$ and $\text{k}r_{di}$ to quadratic terms.
This leads to the following expression for the acoustic velocity:
\begin{eqnarray}
\nonumber 
C_{da}=\frac{eE}{\eta kT_i}r^2_{di}\omega^2_{pd}=\\
=\frac{eE}{\eta kT_i}C^2_{RSY}=C_{RSY}\frac{E}{E_{cr}} \label{eq:cdanew},
\end{eqnarray}
where
\begin{equation}
E_{cr}=\frac{kT_i}{e}\frac{\eta}{C_{RSY}}
\end{equation}
was defined in Ref.~\onlinecite{fink_autowaves_epl} as a critical electric field required for the onset of self-excited wave generation.
\par
According to Eq.~\ref{eq:cdanew}, acoustic speed of dust density waves is proportional to the local electric field $E$.
Simulations \cite{Zobnin2018}, however, show that $E$ can vary about a factor of $2$ over the microparticle cloud drifting in a dc discharge.
Therefore, the latter approach seems to be more promising to provide physical explanation for the observed variation of the phase velocity of the self-excited waves.
Estimations using the microparticle-free values of the plasma parameters (see Sec.~\ref{sec:setup}), $n_d=10^5$~cm$^{-3}$ and $Z_d=2200$ lead to the following numbers: $E_{cr}=1.52$~V/cm, $C_{da}=27.0$~mm/s.
To estimate $C_{da}$, we took the value of the electric field for the microparticle-free positive column, i.e. $E=2.4$~V/cm.  
Assuming the doubled value of $E$ in the head of the cloud as suggested by Ref.~\onlinecite{Zobnin2018}, we obtain the range of $C_{da}$ between $27$ and $54$~mm/s, which is not too far from the experimentally observed range $16$ to $33$~mm/s (see Table~\ref{tab:table2}). 
Since simplified models are used for the determination of $Z_d$ (Eqs.~\ref{Eq:ForceB1}~and~\ref{Eq:ForceB2}) and $C_{da}$, the agreement with the experiment could be considered as fairly well established.
\par
It is notable, that the same approach applied to the growth rate of the dust density waves also gives reasonable estimation.
Growth rate $\gamma$ can be expressed as follows:
\begin{equation}
\gamma=\frac{\omega^2}{\eta C_{RSY}}\frac{E_{cr}}{E}\left(1-\frac{\eta^2}{\omega^2_{pd}}-\frac{E^2_{cr}}{E^2}\right).
\end{equation}
Assuming $\omega=\text{k}V_{ph}$ with $V_{ph}$ taken for region~III (see Table~\ref{tab:table2} and Fig.~\ref{fig:hilbert_result}(c)) and $\text{k}=\text{k}_L=3.9$~rad/mm, we obtain $\gamma$ ranging from $1.06$~mm$^{-1}$ at $E=2.4$~V/cm till $0.96$~mm$^{-1}$ at $E=4.8$~V/cm with the peak value of $1.08$~mm$^{-1}$, reached at $E\approx2.45$~V/cm.
The experimentally measured value of $1.0\pm0.2$~mm$^{-1}$ (Sec.~\ref{sec:wpattern}) is within the estimated interval.
\par

\section{Conclusion}
\label{Sec:Conc}
We presented the results of a microgravity experiment on board the International Space Station performed in the \mbox{PK-4} complex plasma facility, in which the effect of dc discharge polarity reversal on the self-excited dust density waves were studied.
A microparticle cloud was first trapped in a rf discharge and then allowed to drift in a dc discharge with subsequent reversal of the polarity.
At both polarities of the dc discharge, the self-excited dust density waves were observed in a drifting cloud. 
\par
DC plasma, perturbed by a drifting microparticle cloud, exhibited significant non-uniformity, which revealed itself in the plasma emission, microparticle drift velocities (in accord with Ref.~\onlinecite{Zobnin2018}) and also in the self-excited wave pattern.
The phase velocity of the waves in the head of a drifting cloud appeared to be factor of two larger than that in its tail.
Our theoretical estimations (using the acoustic limit of a dispersion relation from Ref.~\onlinecite{usachev_waves}) have shown, that spatial variation of the axial electric field inside a drifting cloud be responsible for the observed variation of the phase velocity.
\par
In spite of the change of the microparticle flow direction after the polarity reversal, the wave continued propagating in the same direction with respect to the camera. 
From the force balance, we found that the gas flow which is responsible for this asymmetry.
\par
After the polarity reversal, the wave pattern exhibited bifurcations.
New wave crests were born in the head of the cloud.
They, however, merged with old crests after $1.5-2$~s of propagation.
We could not explain their appearance, however, the observed growth rate of the newborn crests matched our theoretical estimations. 
\par
The dust density waves are supposed to be excited as a consequence of the ion-streaming instability, associated with the electric field.
We have shown, that the local value of the electric field affects the wave pattern.
However, further investigations (both theoretical and experimental) are obviously necessary to understand the nonlinear physics underlying the bifurcations and better justify the role of the electric field.
\mbox{PK-4} facility offers a unique opportunity to vary the average electric field experienced by the microparticles.
This can be achieved by changing the duty-cycle of polarity switching \cite{pk4rsi}.
This opportunity will be used in a dedicated microgravity experiment, which should give a deeper insight into the problem.

\section{acknowledgement}
All authors greatly acknowledge the joint ESA-Roscosmos ``Experiment Plasmakristall-4'' on board the International Space Station.
We would like to thank Dr. J. Williams and Dr. A. Zobnin for valuable discussions and suggestions.
We also thank Dr. I. Laut for the careful reading of our manuscript.
S. Jaiswal acknowledges the financial support of DLR-DAAD Research Fellowship.
This work was also partially supported by DLR Grants 50WM1441 and 50WM1742.
\section*{References}


\begin{thebibliography}{1}
\bibitem{dawrsy}
N.N. Rao, P.K. Shukla, M.Y. Yu, Planet Space Sci. {\bf 38}, 543 (1990).
\bibitem{barkan_merlino}
A. Barkan, R.L. Merlino, N. D'Angelo, Phys. Plasmas {\bf 2}, 3563 (1995).
\bibitem{merlino25}
R.L. Merlino, J. Plasma Phys. {\bf 80}, 773 (2014).
\bibitem{khrapakag_waves}
V.I. Molotkov, A.P. Nefedov, V.M. Torchinskii, V.E. Fortov, A.G. Khrapak, JETP {\bf 89}, 477 (1999).
\bibitem{ethomas_waves}
E. Thomas, Phys. Plasmas {\bf 13}, 042107 (2006).
\bibitem{usachev_waves}
V.E. Fortov, A.D. Usachev, A.V. Zobnin, V.I. Molotkov, O.F. Petrov, Phys. Plasmas {\bf 10}, 1199 (2003).
\bibitem{piel_waves}
A. Piel, O. Arp, M. Klindworth, A. Melzer, Phys. Rev. E {\bf 77}, 026407 (2008).
\bibitem{pkenefedov_waves}
S.A. Khrapak, D. Samsonov, G. Morfill, H. Thomas, V. Yaroshenko, H. Rothermel, T. Hagl, V. Fortov, A. Nefedov, V. Molotkov, O. Petrov, A. Lipaev, A. Ivanov, Y. Baturin, Phys. Plasmas {\bf 10}, 1 (2003).
\bibitem{schwabe_highlyreswaves}
M. Schwabe, M. Rubin-Zuzic, S. Zhdanov, H.M. Thomas, G.E. Morfill, Phys. Rev. Lett. {\bf 99}, 095002 (2007).
\bibitem{piel_clustering}
K.O. Menzel, O. Arp, A. Piel, Phys. Rev. Lett. {\bf 104}, 235002 (2010).
\bibitem{nosenko_qmachwaves}
V. Nosenko, S.K. Zhdanov, S.-H. Kim, J. Heinrich, R.L. Merlino, G.E. Morfill, Eur. Phys. Lett. {\bf 88}, 65001 (2009).
\bibitem{kortshagen_waves}
U. Kortshagen, Appl. Phys. Lett. {\bf 71}, 208 (1997).
\bibitem{rosenberg_sewaves}
M. Rosenberg, Planet. Space Sci. {\bf 41}, 229 (1993).
\bibitem{du_wagglomeration}
C. Du, H.M. Thomas, A.V. Ivlev, U. Konopka, G.E. Morfill, Phys. Plasmas {\bf 17}, 113710 (2010).
\bibitem{cprmp}
G.E. Morfill, A.V. Ivlev, Rev. Mod. Phys. {\bf 81}, 1353 (2009).
\bibitem{benjamin_daw_pop}
B. Tadsen, F. Greiner, S. Groth, A. Piel, Phys. Plasmas, {\bf{22}}, 113701 (2015).
\bibitem{benjamin_daw_pre}
B. Tadsen, F. Greiner, A. Piel, Phys. Rev. E {\bf 97}, 033203 (2018).
\bibitem{pk4rsi} 
M. Y. Pustylnik, M. A. Fink, V. Nosenko, T. Antonova, T. Hagl, H. M. Thomas, A. V. Zobnin, A. M. Lipaev, A.D. Usachev, V. I. Molotkov, O. F. Petrov, V. E. Fortov, C. Rau, C. Deysenroth, S. Albrecht, M. Kretschmer, M. H. Thoma, G. E. Morfill, R. Seurig, A. Stettner, V. A. Alyamovskaya, A. Orr, E. Kufner, E. G. Lavrenko, G.I. Padalka, E. O. Serova, A. M. Samokutyayev, and S. Christoforetti, Rev. Sci. Instrum., {\bf 87}, 093505 (2016).
\bibitem{chengran_epl}
L. Yang, M. Schwabe, S. Zhdanov, H.M. Thomas, A.M. Lipaev, V.I. Molotkov, V.E. Fortov, J. Zhang, C.  Du, Eur. Phys. Lett., {\bf{117}}, 25001 (2017).
\bibitem{jeremiah1_pre} 
J. D. Williams, Phys. Rev. E, {\bf{89}}, 023105 (2014).
\bibitem{menzel_hilbert_pre}
K. O. Menzel, O. Arp, and A. Piel, Phys. Rev. E, {\bf{83}}, 016402 (2011).
\bibitem{Zobnin2018} 
A.V. Zobnin, A.D. Usachev, O.F. Petrov, V.E. Fortov, M.H. Thoma, M.A. Fink, Phys. Plasmas {\bf 25}, 033702 (2018).
\bibitem{fink_autowaves_epl}
M. A. Fink, S. K. Zhdanov, M. Schwabe, M. H. Thoma, H. H \"ofner, H. M. Thomas and G. E. Morfill, Eur. Phys. Lett., {\bf{102}}, 45001 (2013).
\bibitem{LADAW} 
V.E. Fortov, O.F. Petrov, V.I. Molotkov, M.Y. Poustylnik, V.M. Torchinsky, A.G. Khrapak, A.V. Chernyshev, Phys. Rev. E {\bf 69}, 016402 (2004).
\end{thebibliography}
\end{document}